\documentstyle{aipproc}

\newcommand{\lt}{\linethickness{.9mm}}

\newcommand{\w}{\mbox{\tiny $\wedge$}}
\newcommand{\I}{\mbox{${\bf I}$}}
\newcommand{\spc}{\hspace{.8cm}} 

\title{Notes on black holes and three dimensional gravity }

\author{M\'aximo Ba\~nados}

\address{Departamento de F\'{\i}sica Te\'orica, Universidad de
  Zaragoza,
  Ciudad Universitaria 50009, Zaragoza, Spain.}

\begin{document}

\maketitle

\begin{abstract}
 
These notes are the written version of two lectures delivered at the
VIII Mexican School on Particles and Fields on November 1998. The
level of the notes is basic assuming only some knowledge on
Statistical Mechanics, General Relativity and Yang-Mills theory. After
a brief introduction to the classical and semiclassical aspects of
black holes, we review some relevant results on 2+1 quantum gravity.
These include the Chern-Simons formulation and its affine Kac-Moody
algebra, the asymptotic algebra of Brown and Henneaux, and the
statistical mechanics description of 2+1 black holes. Hopefully, this
contribution will be complementary with the review paper
hep-th/9901148 by the same author, and perhaps, a shortcut to some
recent developments in three dimensional gravity.

\end{abstract}

\section{Introduction} 

During the last three years we have witnessed a rapid progress in the
string theory description of general relativity. Successfull
computations of black hole entropy for extremal and near extremal
solutions \cite{Strominger-V,Callan-M} have made it clear
that the string theory degrees of freedom describes the expected
semiclassical behaviour of general relativity.  This is in sharp
constrast with the more standard approach to quantum gravity either
based on the path integral approach or the Wheeler-de Witt equation
which has provided little information about the fundamental degrees of
freedom giving rise to the Bekenstein-Hawking entropy. In the Loop
representation approach to quantum gravity, a computation of the black
hole entropy has been proposed \cite{Rovelli,Ashtekar-BCK}. However,
in this formulation it is still obscure how to introduce dynamics, and
only the kinematics of spin networks is under control.

In this contribution we shall consider neither string theory nor loop
quantum gravity. Instead, we work in the very simple setting of
three-dimensional quantum gravity whose Lagrangian describes a
well-defined quantum field theory \cite{Achucarro-T,Witten88}. As
motivations to study three-dimensional gravity, let us mention the
following aspects of it. (i) It is a mathematically simple theory
which combines three important branches of physics: General
Relativity, Yang-Mills theory (with a Chern-Simons action), and
two-dimensional Conformal Field Theory. (ii) The mathematical tools
are surprinsingly similar to those used in string theory, with a
centrally extended Virasoro algebra \cite{BH} as one of its main
ingredients. (iii) The space of solutions contains particle solutions
\cite{Deser-JT} and black holes \cite{BHTZ}, thus making it
interesting from the dynamical point of view. 

In these notes, we shall mainly be interested in quantum black holes
in three dimensions. Our goal is to give, in a somehow self-contained
way, a derivation of Strominger's \cite{Strominger97} proposal for the
statistical mechanical origin of the three-dimensional black hole
entropy. We refer the reader to
\cite{Martinec98,Henningson-S,Giveon-KS,deBoer-ORT} for the stringy
aspects of Strominger's result. See also \cite{Skenderis99} for a
recent review. We shall concentrate here on the gravitational aspects.
For a detailed and complete treatment of three dimensional gravity we
refer to the recent book by Carlip \cite{Carlip-book}. 

In Sec. \ref{Sch} we shall briefly review, at the most basic level,
some of the main properties of the Schwarzschild solution, as well as
the three-dimensional black hole \cite{BHTZ}. In Sec. \ref{CS} we
review the Chern-Simons formulation of three-dimensional gravity.
Particular empahsis is given to the absence of bulk degrees of
freedom, and a quick derivation of the affine Kac-Moody algebra is
presented. Finally, in Sec. \ref{Entropy} we derive the Brown-Henneaux
conformal algebra, and its statistical mechanical
\cite{Strominger97,Carlip98-1}
implications.

\section{Classical and semiclassical black holes} 
\label{Sch}

\subsection{The black hole spacetime} 

The Schwarzschild metric $(r>r_0)$,
\begin{equation}
ds^2 = -(1-2M/r)dt^2 + (1-2M/r)^{-1} dr^2 + r^2 d\Omega^2, 
\label{sch}
\end{equation}
is an exact solution of the Einstein vacuum equations
\begin{equation}
G_{\mu\nu} = R_{\mu\nu} - {1\over 2} g_{\mu\nu} R =0, 
\label{EE}
\end{equation}
representing the geometry outside a collapsing star of mass $M$ and
radius $r_0$. One of the most surprising predictions of General
Relativity, which caused much confusion in the past, is the
appearance of a singularity in the metric for the particular value of
$r$: 
\begin{equation}
r=:r_+ = 2M  \ \ \ \ \ \mbox{(Event Horizon)}.
\label{r+}
\end{equation}
If the radius of the star $r_0$ is less that $r_+$ then the solution
(\ref{sch}), which is valid for $r>r_0$, has a singularity at $r=r_+$.
Furthermore, in the region $0<r<r_+$ where the metric is
again regular, $r$ is a timelike coordinate while $t$ is spacelike.
Finally, at $r=0$ the curvature blows up making gravitational forces
divergent there. This means, in particular, that no observer can reach
the singularity without being destroyed.  The possibility of making
experiments near the singularity is prevented by another fact: any
observer that crosses the event horizon $r=r_+$ will never come back,
at least not according to the classical Einstein equations. We shall
prove this below. Quantum mechanically, particles can tunnel out of
the black hole and escape to infinity. This is Hawking's famous
discovery of black hole radiation \cite{Hawking74}. However, according
to Hawking's description there is no correlation between the particles
that fall into the black hole with the ones that escape. This point is
actually a matter of discussion and there is no agreement yet. We will
not have time here to describe in any detail this very interesting
work. We refer the interested reader, for example, to \cite{Page93}
for a review with an extensive list of references.  

Let us now briefly show how to deal with the $r=r_+$ singularity in
(\ref{sch}). This will allow us to see why observers cannot travel
back once they have crossed the horizon. We shall also infer the
value of Hawking's temperature via a geometrical argument. 

The analysis that follows does not depend on the details of the
Schwarzschild solution but only on some general properties of black
holes. We consider general metrics in $d$ dimensions of the form, 
\begin{equation}
ds^2 = -f(r) dt^2 + f^{-1}(r) dr^2 + ds_{d-2}^2,
\label{bh/f}
\end{equation}
where $ds_{d-2}^2$ represents the metric of a $S_{d-2}$ sphere, or
some other compact or non-compact surface.   The function $f(r)$
satisfies the following two properties: (i) There exists a value of
$r$ denoted as $r_+$ such that $f(r_+)=0$; (ii) The derivative of $f$
at $r_+$ is different from zero,
\begin{equation}
\alpha \equiv {1 \over 2} \left. { df(r) \over dr } \right|_{r=r_+}, \
\ \ \ \ \ \ \ \ \  \alpha \neq 0.
\label{alpha}
\end{equation}

Most known (non-extreme!) black holes have a metric of this form, or
at least there is a plane on which the metric near $r=r_+$ looks like
the first two terms in (\ref{bh/f}). The extreme black holes do not
fall into the above class of metrics because the function $f(r)$ has a
second order zero and thus $\alpha=0$. These black holes play an
important role in string theory because they are related to BPS
states.

For the Schwarzschild black hole, the function $f$ is given by $f(r)=
1-2M/r$ and $f'(r_+) = 1/2M$. This means that $\alpha=1/4M$ which is
indeed different from zero. Other examples are: The Reissner-Nordstrom
black hole with $f(r) = 1-2M/r + e^2/r^2$ and $e$ is the electric
charge. In this case, $\alpha$ is different from zero provided $M\neq
e$; The 2+1 black hole (to be studied in detail in the next section)
with $f(r) = -M + r^2/l^2$ and $l$ is related to a cosmological
constant; The d-dimensional Schwarzschild solution with $f(r) =
1-2M/r^{d-3}$; plus all the (non-extreme) stringy black holes
\cite{Horowitz92}, as well as other higher dimensional
situations\cite{BTZ3}.  Students are encouraged to compute the value
of $\alpha$ for each of these black holes, as we shall see soon, this
number is essentially Hawking's temperature for each of these objects. 

The metric (\ref{bh/f}) is singular at the event horizon, just as the
Schwarzschild metric is. To cure this singularity we introduce the
following new set of coordinates. We change $\{r,t\}$ to $\{u,v\}$
according to,
\begin{eqnarray}
u = g(r) \cosh \alpha t,  \nonumber\\ 
v = g(r) \sinh \alpha t, 
\label{uv}
\end{eqnarray}
where the function $g(r)$ is defined by, 
\begin{equation}
g(r) = \exp\left( \alpha \int^r {dr'\over f(r')} \right). 
\label{g}
\end{equation}
This change of coordinates has the following properties. The event
horizon $r=r_+$ is mapped into the lines $u=\pm v$.  The metric in
terms of $u,v$ reads, 
\begin{equation}
ds^2=\Omega^2(r) (-dv^2 + du^2) + r^2 d\Omega^2, 
\label{kru/f}
\end{equation}
where the function, 
\begin{equation}
\Omega^2(r) = {f(r) \over \alpha^2 g^2(r) },  
\label{Omega}
\end{equation}
is regular at $r=r_+$. The regularity of $\Omega^2$ holds provided
$f(r)$ has a simple pole at $r=r_+$.  It is easy to see using
L'Hopital rule that the zero in $f(r)$ is cancelled by $g^2(r)$
provided $\alpha$ is chosen as in (\ref{alpha}). Note that the above
coordinate change does not depend on the details of the function
$f(r)$, provided it has a single pole at $r=r_+$. Of course, our
formulae for the conformal factor and change of coordinates reproduce
the usual expressions when applied to particular situations like the
Schwarzschild black hole (see \cite{Hawking-E}). 

The coordinates $u$ and $v$ are called Kruskal coordinates and their
range is $-\infty <u,v<\infty$. These coordinates can be compactified
(see \cite{Hawking-E} for more details on this points) and led to the
Penrose diagram shown in Fig. 1\footnote{The Penrose diagram shown
here assumes asymptotic flatness. This means that the function $f(r)$
satisfies $f(r\rightarrow \infty) \rightarrow 1$. This is not the case
for asymptotically anti-de Sitter black holes for which $f\rightarrow
r^2$. The Penrose diagram in that case can still be drawn and differs
only in the asymptotic structure, not the properties surrounding the
horizon. See second Ref. in \cite{BHTZ}.}.  Region $\I$ is the black
hole
exterior $r>r_+$ and region $\I\I$ its interior ($r<r_+$). It should
be clear from the figure that an observer situated in region $\I\I$
cannot go back to region $\I$ because he or she would need to travel
faster than light. The fate of any future-directed (timelike) observer
is to hit the singularity. 

\setlength{\unitlength}{.91mm}
\begin{center}
\begin{picture}(150,50)
\footnotesize
\put(60,10){\lt\line(1,0){30}}
\put(60,40){\lt\line(1,0){30}}
\put(45,25){\lt\line(1,1){15}}
\put(90,10){\lt\line(1,1){15}}
\put(60,10){\lt\line(1,1){30}}
\put(45,25){\lt\line(1,-1){15}}
\put(90,40){\lt\line(1,-1){15}}
\put(60,40){\lt\line(1,-1){30}}
\put(75,7){\makebox(0,0){$r=0$}}
\put(75,43){\makebox(0,0){$r=0$}}
\put(84,34){\makebox(0,0){$r=r_+$}}
\put(84,16){\makebox(0,0){$r=r_+$}}
\put(99,34){\makebox(0,0){$r=\infty$}}
\put(99,16){\makebox(0,0){$r=\infty$}}
\put(65,25){\makebox(0,0){\I'}}
\put(85,25){\makebox(0,0){\I}}
\put(75,31){\makebox(0,0){\I\I}}
\put(75,19){\makebox(0,0){\I\I'}}
\end{picture}
Fig. 1 \\
Penrose diagram for a Schwarzschild black hole. 

\end{center}

\subsection{Semiclassical black holes. The Gibbons-Hawking
approximation} \label{Gibbons-H}

The combination of Euclidean field theory together with the
coordinate change (\ref{uv}) suggest in a very direct way that black
holes should have a
non-zero temperature. The Euclidean formalism (sometimes called the
Euclidean sector) is obtained by setting $\tau = it$, and the metric
(\ref{sch}) becomes Euclidean. Consider again the change of
coordinates (\ref{uv}) in the Euclidean formalism. The hyperbolic
functions will be replaced by their trigonometrical versions and it is
clear that the Euclidean time variable needs to be an angle $0\leq
\alpha\tau <2\pi$. In the Euclidean sector, and near the horizon, the
change (\ref{uv}) is nothing but the relation between polar and
cartesian coodinates in $\Re^2$. In fact, the topology of the
Euclidean Schwarzschild black hole is $\Re^2\times S_2$ where the
origin of $\Re^2$ is situated at the horizon $r=r_+$.  The Euclidean
sector does not see the inner region of the black hole $r<r_+$.  

Following the usual practice of Euclidean field theory we define
the inverse temperature as the Euclidean time period $(\hbar=1)$, 
\begin{equation}
\beta = {2\pi \over \alpha}. 
\label{T}
\end{equation}
So far this is only a mathematical trick with no real physics meaning.
However, it turns out that the temperature $T=1/\beta$ defined in
(\ref{T}) coincides exactly with Hawking's evaporation temperature.
For the Schwarzschild black hole, we recall that $\alpha= 1/4M$, 
this yields the famous Hawking result,
\begin{equation}
T_H = {1 \over \beta_H} = {1 \over 8\pi M}. 
\label{TH}
\end{equation}
Now, integrating the first law $dM = T dS$ we find the 
Bekenstein-Hawking formula for black hole entropy, 
\begin{equation}
S = {A \over 4},
\label{S}
\end{equation}
where $A=4\pi r_+^2 $ is the area of the event horizon ($r_+=2M$). 

For our purposes, this ``derivation" of the black hole temperature and
entropy has an important meaning: geometry knows that black holes
radiates. In other words, the very deep origin of Hawking's process is
not contained only on the matter fields surrounding a black hole but
rather on the gravitational (perhaps string) degrees of freedom.  This
point of view is further supported by the Gibbons-Hawking calculation
of the Schwarzschild black hole partition function which we now
describe.

Let us briefly review here the results presented in \cite{Gibbons-H}
in the simplest case of a non-rotating black hole. Our main tool 
will be again the analogy between Euclidean field theory and
statistical mechanics. 

Consider the functional integral,
\begin{equation}
Z[h] = \int Dg\, e^{-I[g,h]},
\label{Z}
\end{equation}
where,
\begin{equation}
I[g,h] = -\frac{1}{16\pi G}\left(\int_M \sqrt{g} R + 2\int_{\partial
M} \sqrt{h} K \right).
\label{I0}
\end{equation}
is the Euclidean gravitational action appropriated to fix the metric
at the boundary. Here $h_{ij}$ is the 3-metric induced on $\partial
M$. The boundary term is added to the action to ensure that $I$ has an
extremum when $h$ is fixed. $Dg$ denotes the sum, modulo
diffeomorphisms, over all metrics with $h_{ij}$ fixed.  As it is well
known, the formula (\ref{Z}) is purely formal and cannot be given a
precise mathematical meaning. This is because gravity is not
renormalizable and the perturbation expansion for (\ref{Z}) is not
well-defined. To make things worst, the action (\ref{I0}) is not
bounded from below/above, not even in the Euclidean formulation.  It
is possible to find sequences of Euclidean manifolds $M_i$ for which
the value of the action (\ref{I0}) goes to minus/plus
infinity\cite{Gibbons77}. 

Although (\ref{Z}) cannot be computed in general, its saddle point
approximation around some classical solutions gives interesting
results. Incidentally, we mention here that the evaluation of the
action $I$ on classical solutions has become crucial in the recently
discovered adS/CFT correspondence
\cite{Maldacena97,Gubser-KP,Witten98}. The first example of an
evaluation of (\ref{Z}) was performed by Gibbons and Hawking
\cite{Gibbons-H} who considered the Euclidean Schwarzschild black hole
(\ref{sch}) with mass $M$. The mass $M$ and Euclidean period $\beta$
are related by (\ref{TH}) in order to avoid singularities (sources) in
the Euclidean metric.  The value \footnote{Actually, the value of $I$
diverges and needs to be regularized. See \cite{Gibbons-H} for details
and \cite{Brown-,BTZ4} for an  evaluation of $I$ using Hamiltonian
methods on which the regularization is automatic.} of $Z$ in the
saddle point (\ref{sch}) is 
\begin{equation}
Z[\beta] \sim e^{-\beta^2/16\pi}.
\label{Zsch}
\end{equation}
This result is quite remarkable. The thermodynamical formula for the
average energy $M = -\partial\log Z/\partial \beta$ reproduces
(\ref{TH}) and confirms that $\beta$ is the inverse temperature of the
black hole.  In the same way, the average entropy $S=\ln Z - \beta
\partial_\beta \ln Z$ reproduces (\ref{S}). This result confirms once
again that the black hole thermal properties are present in a pure
quantum theory of gravity, and not only in the interaction of a
classical background metric with quantized fields.

Hawking's discovery of black hole evaporation is one of the most
important results in the theory of general relativity and quantum
mechanics.  We refer the reader, for example, to the classic books by
Birrell and Davis \cite{Birrell-D} and Wald \cite{Wald84} for a
detailed discussion on quantum black holes, and in particular, quantum
field theory on curved spacetimes. These books were written before 
black holes became important in string theory. See \cite{Horowitz96}
for a review on the string theory approach to black holes.  

We shall now depart from the Schwarzschild four dimensional black hole
and go down to three dimensions where a black hole solution exists
\cite{BHTZ} having many of the features of the Schwarzschild solution,
but it is far simpler mathematically.  

Consider the action for three-dimensional gravity with a negative
cosmological constant $\Lambda=-2/l^2$,
\begin{equation}
I= {1 \over 16\pi G} \int \sqrt{-g} \left(R + {2 \over l^2}
\right)dx^3.
\label{I3}
\end{equation}
In three dimensions it is convenient to keep the fundamental constants
because, due to the cosmological constant, there are two fundamental
length parameters: Plank's length $l_p=\hbar G$ and the cosmological
radius $l$.  

The equations of motion derived from this action are solved by 
the (non-rotating) three-dimensional black hole\cite{BHTZ}
\begin{equation}
ds^2 = -\left(-8MG + {r^2 \over l^2}\right) dt^2 +  
\left(-8MG + {r^2 \over l^2}\right)^{-1} dr^2 + r^2 d\varphi^2.
\label{BHTZ}
\end{equation}
Angular momentum as well as electric charge can be added easily, see
\cite{BHTZ}.  

As for the Schwarzschild metric, we can go to the Euclidean sector and
discover that the time coordinate is periodic. The associated
temperature is, 
\begin{equation}
T_3 = {\sqrt{M} \over 2\pi l^2},
\label{T3}
\end{equation}
and the entropy is again given by (\ref{S}), but now $A=2\pi r_+$ is
the perimeter length of the horizon.

A word of caution is necessary here. Contrary to the Schwarzschild
case, the metric (\ref{BHTZ}) is not asymptotically flat. This means
that the Euclidean period cannot be defined as the proper length of
the time coordinate at infinity. Note that the limit $r\rightarrow
\infty$ of the Euclidean  Schwarzschild metric yields a well defined
metric at the ``boundary" (infinity is not really a boundary) with the
topology $S_1 \times S_2$. $S_1$ corresponds to the periodic time
coordinate, while $S_2$ to the angular sphere. The limit $r\rightarrow
\infty$ of (\ref{BHTZ}) is not well defined. At infinity, one can only
define a conformal class of metrics . This is a three-dimensional
example of the adS/CFT correspondence
\cite{Maldacena97,Gubser-KP,Witten98} first studied in \cite{BH}. A
more rigorous definition for the temperature can be given by noticing
that the topology of (\ref{BHTZ}) in the Euclidean sector is a solid
torus. The temperature is related to the complex structure of the
torus\cite{BBO,Maldacena-S} by,
\begin{equation}
\tau = {\beta \over 2\pi} \left( \Omega + {i \over l} \right), 
\label{tau}
\end{equation}
where $\beta$ is the period of the Euclidean time coordinate, 
and $\Omega$ is the angular velocity. In the non-rotating case,
$\Omega=0$.     

As before, one can write down the three-dimensional partition function
in the saddle point approximation provided by the solution
(\ref{BHTZ}). This yields\cite{BHTZ} (see also \cite{BM} for a
Lagrangian approach),
\begin{equation}
Z_3 \sim e^{\pi^2 l^2/(2G\beta) }, 
\label{Z3}
\end{equation}
and it is direct to check that the thermodynamical formulae for the
average energy and entropy is consistent with (\ref{T3}). 

Our main motivation to study three dimensional quantum gravity is to
try to give a precise meaning to the formula (\ref{Z}) in three
dimensions. In other words, we hope that in three dimensions (\ref{Z})
could be well-defined mathematically, and provide the semiclassical
limit (\ref{Z3}). If this is true, then one should be able to
extract from the exact formula for $Z$ which are the degrees of
freedom giving rise to the black hole entropy. 

The main mathematical device that we shall use is the
Chern-Simons formulation of three dimensional gravity
\cite{Achucarro-T,Witten88}.  This formulation makes manifest the
fact that three dimensional gravity does not have any bulk degrees of
freedom and it is renormalizable. Still, this does not mean that the
problem is trivial because the relevant group, see below, is $SL(2,C)$
which is not compact.  The evaluation of the partition
function for the black hole problem using the Chern-Simons formulation
was initiated by Carlip \cite{Carlip97}. Some clarification on the
boundary conditions and the role of ensembles can be found in
\cite{BBO}. A string theory approach can be found in 
\cite{Maldacena-S}. Further developments on the modular properties of
the partition function in three dimensions have recently appeared in
\cite{Brotz-OR99}.

\section{2+1 gravity as a Chern-Simons theory.}
\label{CS}

\subsection{First order form of the Euclidean action}

We start with the (Palatini) Euclidean action for three dimensional
gravity with a negative cosmological constant $\Lambda = -2/l^2$,
\begin{equation}
I[g_{\mu\nu},\Gamma^\rho_{\mu\nu}] = {1 \over 16\pi G} \int \sqrt{g}
\left( g^{\mu\nu}R_{\mu\nu}(\Gamma) +
{2 \over l^2}
\right)
\label{Ig}
\end{equation}
The discovery of Ach\'ucarro and Townsend is that in three dimensions
one can replace the metric by two Yang-Mills fields such that both the
structure of the action and equations of motion simplifies
enormously.   

This is achieved in various steps. First, we use the Palatini
formalism. The idea of this formalism is to note that the Ricci tensor
$R_{\mu\nu}$ depends on the metric only through the Christoffel symbol
$\Gamma^\rho_{\mu\nu}$. Then, it follows that if one treats
$g_{\mu\nu}$ and $\Gamma^\rho_{\mu\nu}$ as independent variables in
the action (\ref{Ig}), the equations of motion yield the expected
relation $g_{\mu\nu;\lambda}=0$ between the metric and connection.
Next, we make a change of coordinates from the coordinate basis
$\partial_\mu$ to orthonormal coordinates on which the metric is flat.
The matrix that makes this change is called the triad and is defined
by the formula,
\begin{equation}
g_{\mu\nu} = e^a_\mu\, \eta_{ab}\, e^b_\nu
\label{e}
\end{equation}
Clearly, $e^a_\mu$ is defined only up to a (local) Lorentz rotation
because if $\Lambda$ is an element of the Lorentz group then, by
definition, $\Lambda \eta \Lambda^{-1} = \eta$.  Equation (\ref{e}) is
nothing but the transformation of a tensor under a change of
coordinates described by the matrix $e^a_\mu$. We also need to
transform the Christoffel symbol which is not a tensor but we know
its transformation law under $e^a_\mu$, 
\begin{equation}
\Gamma^\sigma_{\mu\nu} = e_a^\sigma
\omega^a_{\ b\nu} e^b_\mu + e^\sigma_a e^a_{\mu,\nu}
\label{w}
\end{equation}
where $w^{ab}_\mu$, known for historical reasons as the spin
connection, is the new `Christoffel symbol' in the new coordinates.
Eq. (\ref{w}) is often written in the literature as  $e^a_{\mu;\nu}=0$
where the semicolon denoted full covariant derivative, or as $D_\mu
e^a_\nu = \Gamma^\rho_{\nu\mu} e^a_\rho$ where $D_\mu$ denotes
covariant derivative in the spin connection.  These formulae are, of
course, equivalent to (\ref{w}). Note that in (\ref{w}) we have only
transform two indices. The reason is that the Christoffel symbol is a
1-form connection for the group $GL(4,\Re)$, $\Gamma^\mu_{\
\nu}=\Gamma^\mu_{\ \nu\rho}dx^\rho$. The next object we would
like to write in the new coordinates is the curvature tensor. The
curvature tensor is a tensorial 2-form, for that reason we only
transform two of its four indices as, 
\begin{equation}
R^{\lambda\sigma}_{\ \ \mu\nu} = e^\lambda_a e^\sigma_b 
R^{ab}_{\ \ \mu\nu}
\label{R}
\end{equation}
where $R^{ab} = dw^{ab} + w^a_{\ c} \w w^{cb}$.  With formulae
(\ref{e}), (\ref{w}) and (\ref{R}) at hand we can
prove the identity
\begin{equation}
 \int \epsilon_{abc} R^{ab} \w e^c = \int \sqrt{g} R
\label{Ige}
\end{equation}
The relevant steps are (we go from left to right),
\begin{eqnarray}
\int \epsilon_{abc} R^{ab} \w e^c &=& \int
\epsilon^{\mu\nu\lambda}\epsilon_{abc} \left(\frac{1}{2}
R^{ab}_{\ \ \mu\nu}\right) e^c_\lambda \label{um} \nonumber\\
&=& \frac{1}{2}\int 
\epsilon^{\mu\nu\lambda}\epsilon_{abc} R^{\alpha\beta}_{\ \ \mu\nu} 
e^a_\alpha e^b_\beta e^c_\lambda \nonumber\\
&=& \frac{1}{2}\int 
\epsilon^{\mu\nu\lambda} \epsilon_{\alpha\beta\lambda}\, e \,
R^{\alpha\beta}_{\ \ \mu\nu} .
\label{Ige/}
\end{eqnarray}
In the second line we have used (\ref{R}), and in the third line
$\epsilon_{abc} e^a_\alpha e^b_\beta
e^c_\lambda=e\epsilon_{\alpha\beta\lambda}$ with $e$ equal to the
determinant of $e^a_\mu$, and $\epsilon^{\mu\nu\lambda}
\epsilon_{\alpha\beta\lambda} =\delta^{[\mu\nu]}_{[\alpha\beta]}$
(recall that we are working in the Euclidean formalism).  It
should be clear that the last line in (\ref{Ige/}) is equal to the
right hand side of (\ref{Ige}).

Collecting all formulae together we arrive at the new action for
three-dimensional gravity,
\begin{equation}
I[e^a,w^{ab}] = {1 \over 16\pi G} \int \epsilon_{abc} 
\left( R^{ab} + {1 \over 3l^2} e^a\w e^b \right)\w e^c .
\label{Iew}
\end{equation}
The action $I[e,w]$ is equal to the action $I[g,\Gamma]$ shown in
(\ref{Ig}). Besides notation issues there is a conceptual consequence.
The action  (\ref{Iew}) is perfectly well defined even is
the metric is degenerate.  In this sense, the triad formulation
provides a generalization for the Einstein-Hilbert action.  

The last step before we can write the Chern-Simons action is to define
the new spin connection\footnote{With this definition of $w^a$
the torsion becomes $T^a=de^a + \epsilon^a_{\ bc} w^b \wedge e^c$.}
$w^a$ and curvature $R^a$, 
\begin{equation} 
w^a = -(1/2) \epsilon^a_{\ bc} w^{bc}, \ \ \ \ \ \ \ 
R^a = -(1/2) \epsilon^a_{\ bc} R^{bc}
\end{equation}
with $R^a = dw^a + \frac{1}{2}\epsilon^a_{\ bc} w^b\w w^c$.

We are now ready to make the connection with Chern-Simons theory.
Let $x$ be a complex number and let $A^a$ and $\bar A^a$ to fields
related to $e$ and $w$ by,
\begin{equation}
A^a = w^a + x e^a, \ \ \ \ \bar A^a = w^a - x e^a.
\label{Awe}
\end{equation}
The relation between Chern-Simons theory and three dimensional general
relativity follows from the equality:
\begin{eqnarray}
2 e_a R^a + {x^2 \over 3} \epsilon_{abc} e^a e^b e^c 
&= &\ {1 \over 2x} ( A_a dA^a + {1\over 3} \epsilon_{abc} A^a A^b A^c)
\nonumber\\
 && - {1 \over 2x} (\bar A_a d \bar A^a + {1\over 3} \epsilon_{abc}
 \bar A^a \bar
 A^b \bar A^c ) + dB.
\label{CS0}
\end{eqnarray}
This relation is true regardless the signature of spacetime or sign of
the cosmological constant.  Just plug  (\ref{Awe}) into the right hand 
side of (\ref{CS0}) and obtain the left hand side. ($dB$ is a total
derivative term.) 

Depending on the signature of spacetime and cosmological constant $x$
need to be complex or real. We shall be interested here in the
Euclidean gravity with a negative cosmological constant. In the case,
$x$ is purely imaginary.

\subsection{Chern-Simons action}
\label{con}

From the equality (\ref{CS0}) with $x=i/l$ it follows that the
Einstein-Hilbert  action (\ref{Ig}) for Euclidean three dimensional
gravity can be written in the form,
\begin{equation}
I[g,\Gamma] = i I[A]- i I[\bar A],
\label{IAA}
\end{equation}
where $I[A]$ is the Chern-Simons action,
\begin{equation}
I[A] = \frac{k}{4\pi} \int \mbox{Tr} (A dA + \frac{2}{3} A^3),
\label{ICS}
\end{equation}
at level\footnote{The sign of $k$ depends on the identity
$\sqrt{g}=\pm e$ where $e$ is the determinant of the triad. This sign
determines the relative orientation of the coordinate and orthonormal
basis. We have chosen here the plus sign which means that we work with
$e>0$.},
\begin{equation}
k =- {l \over 4G}.
\label{k}
\end{equation}
In (\ref{IAA}) we have defined 
\begin{equation}
 A=A^aJ_a, \quad  \bar A = \bar A^a J_a.
\label{AbarA}
\end{equation}
where the $SU(2)$ generators are given by,  
\begin{equation}
J_1 = {i\over 2} \left( \begin{array}{cc}   0 &  1  \\
                            1 &  0   \end{array} \right), \ \ \
J_2 = {1\over 2} \left( \begin{array}{cc}   0 &  -1  \\
                           1 &  0   \end{array} \right), \ \ \
_3 = {i\over 2} \left( \begin{array}{cc}   1 &  0  \\
                            0 &  -1   \end{array} \right),
\label{conventions}							
\end{equation}
and satisfy $[J_a,J_b]=\epsilon_{ab}^{\ \ c} J_c$,  Tr$(J_a J_b) 
= -(1/2) \delta_{ab}$. Note that $\bar A$ is not the complex conjugate
of $A$. 

Let $F^a$ and $\bar F^a$ the Yang-Mills curvatures associated to $A^a$
and $\bar A^a$. From the point of view of the equations of motion, the
relation between Chern-Simons theory and general relativity is
contained in the fact that the Chern-Simons equations,  
\begin{equation}
F^a =0, \ \ \ \ \ \bar F^a=0,  
\label{EqCS}
\end{equation}
are equivalent to the three-dimensional Einstein equations.
Thus, studying the space of solutions of (\ref{EqCS}) we are studying
general relativity.  

The 1-form $A^a$ is a $SL(2,C)$ Yang-Mills gauge field because in
(\ref{Awe}) $x$ is imaginary.  For Minkowskian gravity $x=1/l$ is real
and the relevant group is $SO(2,1)\times SO(2,1)$.

\subsection{Chern-Simons dynamics. Kac-Moody symmetry} 

Once we have proved the equality between the Chern-Simons and
gravitational actions we can forget about metrics and work with
Yang-Mills fields which are much more tractable. We should keep in
mind however that the Chern-Simons action is a generalization to
general relativity in the sense that it can accept degenerate metrics.  

The classical dynamics of Chern-Simons theory is simple to analyse. 
First, we note that the Chern-Simons action is already in Hamiltonian
form. In the 2+1 decomposition of the gauge field $A^a = A^a_0 dt +
A^a_i dx^i$, the Chern-Simons action reads,
\begin{equation}
I[A_i,A_0] = \frac{k}{8\pi} \int dt \int_\Sigma  \epsilon^{ij}
\delta_{ab} (A^a_i \dot A_i^b - A^a_0 F^b_{ij} ).
\label{ICSH}
\end{equation}
The coordinates $x^i$ are local coordinates on the spatial surface
denoted by $\Sigma$. This action has $2N$ dynamical fields $A^a_i$
($a=1,...,N$; $i=1,2$) and $N$ Lagrange multipliers $A^a_0$. The
dynamical fields satisfy the basic equal-time Poisson bracket
algebra,
\begin{equation}
\{A^a_i(x),A^b_j(y) \} = \frac{4\pi}{k} \epsilon_{ij} \delta^{ab}
\delta^2(x,y).
\label{pAA}
\end{equation}
The equation of motion with respect to $A_0$ leads to the constraint
equation,
\begin{equation}
{k \over 8\pi} \epsilon^{ij} F^a_{ij} \approx 0.  
\label{G=0}
\end{equation}
which (after properly taken into account boundary condition and
boundary terms if the spatial surface has a boundary) generates the
gauge transformations $\delta A^a_i=D_i\lambda^a$ in the Poisson
bracket (\ref{pAA}).  

Because the equations of motion of Chern-Simons theory are $F=0$ we
know that there are no local degrees of freedom in this theory. It is
instructive however to count the number of degrees of freedom per
point using the Dirac formalism. We have $2N$ dynamical variables
subject to $N$ constrains. These constraints are of first class and
generate the $N$ local gauge transformations. Thus the total
number of local degrees of freedom is indeed zero. This does not
mean that the action is trivial.  There are an infinite number of
degrees of freedom associated to the breakdown of gauge invariance at
the boundary, plus a finite number associated to holonomies along
non-contractible loops.  Here we shall not consider the holonomies. We
refer the reader to \cite{Witten88,Nelson-R}. 

The boundary degrees of freedom in Chern-Simons theory can be
understood in many different ways.  Their existence was first
indicated in \cite{Witten89}, and Carlip \cite{Carlip95,Carlip97}
first pointed out that they may be responsible for the
three-dimensional black hole entropy. 

These degrees of freedom are somehow a matter of interpretation
rather than a specific calculation.  The point is that, at the
boundary, is incorrect to identify configurations that differ by a
gauge transformation. As discussed in \cite{Carlip95} this follows
from boundary terms arising in the transformation of the Chern-Simons
action under gauge transformations.  Alternatively, following
\cite{Regge-T}, one can see that at the boundary the transformations
$\delta A^a_i=D_i\lambda^a$ are not generated by constraints and
therefore they do not represent proper gauge transformations. In
summary, the symmetry is still there but its interpretation is
different. 

This point can be exhibited in the following calculation. This  
analysis is taken from \cite{Moore-S} and \cite{Witten84},
with minor modifications.  To simplify the notation, let us use
differential form notation in the spatial manifold $A=A_idx^i$. We
rewrite the action (\ref{ICSH}) in the form, 
\begin{equation}
I[A,A_0] = \frac{k}{4\pi} \int dt \int_\Sigma (A \w \dot A-A_0 F),
\label{ICSH2}
\end{equation}
where the symbol $\int$ includes the trace Tr. The constraint $F=0$
implies,
\begin{equation}
A = g^{-1} d g, 
\label{Ag}
\end{equation}
from where we derive two useful identities, 
\begin{equation}
\delta A = D(g^{-1}\delta g ), \spc  \dot A = D (g^{-1}\dot g ).
\label{dAg}
\end{equation}
$D$ represents the covariant derivative in the flat connection $A$
given in (\ref{Ag}): $D = d + [A,~]$. Our goal is to compute the
commutator of two solutions of the form (\ref{Ag}). 

Consider a non-canonical Lagrangian of the form $L = l_a(z) \dot z^a$
whose variation reads 
\begin{equation}
\delta L = \delta z^a \sigma_{ab} \dot z^b, \ \ \ \ \ \ \ 
\sigma_{ab}(z) = \partial_a l_b - \partial_b l_a.
\label{ls}
\end{equation}
If $\sigma$ is non-degenerate, the Poisson bracket of $z^a$ with
itself is given by
\begin{equation}
\{z^a,z^b\} = J^{ab}(z)
\label{WP}
\end{equation}
where $J$ is the inverse of $\sigma$, $J^{ab} \sigma_{bc}
=\delta^a_c$. The Jacobi identity for $J$ follows from the Bianchi
identity for $\sigma$. If $L =  p \dot q$ the above construction
yields $[q,p]=1$, as expected.  Following \cite{Witten84}, we shall
use this method to compute the Poisson bracket between solutions of
the form (\ref{Ag}). 

The idea is to replace the solution (\ref{Ag}) in the action
(\ref{ICSH2}) and compute its variation on the surface (\ref{Ag}).
Since after replacing (\ref{Ag}) in (\ref{ICSH2}) only the kinetic
term survives the variation of $I$ reads, 
\begin{eqnarray}
\delta I &=& -\frac{k}{2\pi} \int \dot A \w \delta A, \nonumber \\
&=&-\frac{k}{2\pi}\int_\Sigma D(g^{-1}\dot g ) \w D(g^{-1}\delta g
),\nonumber\\ 
&=& \frac{k}{2\pi}\int_{\partial\Sigma}D_\varphi(g^{-1}\dot g)\,
g^{-1}\delta g . 
\end{eqnarray}
The last equality follows from $D\w D=F=0$. We thus find that the
variation of $I$ on the surface (\ref{Ag}) depends only on the
boundary values of $g$.  This is of course the well known fact that
the variation of the $WZW$ action can be written as a local functional
of the boundary. It also means that the only non-trivial degrees of
freedom arise at the boundary, and they are the values of $g$ at the
boundary. Using (\ref{dAg}), the variation of $I$ can be written as,
\begin{equation}
\delta I = \frac{k}{2\pi} \int_{\partial\Sigma} 
\dot A_\varphi \, {1 \over D_\varphi} \, \delta A_\varphi
\end{equation}
where $1/D_\varphi$ is the inverse of the operator $D_\varphi =
\partial_\varphi + [A_\varphi,\ ]$ which
we assume exists (we exclude functions satisfying $D_\varphi f=0$).
Comparing this variation with (\ref{ls}) and
(\ref{WP}), we find the Poisson bracket of $A_\varphi$ with itself,  
\begin{eqnarray}
\{A_\varphi,A_\varphi \} = \frac{2 \pi}{k} D_\varphi 
\label{KM'}
\end{eqnarray}
where the derivative term in $D_\varphi$ should be understood as the
derivative of a Dirac delta function. Finally, we make a Fourier
expansion,
\begin{equation}
A(\varphi)  = \frac{2}{k} \sum_n T^a_n e^{in\varphi}
\end{equation}
and obtain the quantum commutator ($\hbar=1$)
\begin{equation}
[ T^a_n , T^b_m] = i\epsilon^{ab}_{\ \ c} T^c_{n+m} + 
n \frac{k}{2} \delta^{ab} \delta_{n+m}  
\label{KM}
\end{equation}

Some comments are in order here. 

(i) It is clear that the equations of motion do not force $A_\varphi$
to be zero. Actually, in the sector with chiral boundary conditions
$A_\varphi$ is arbitrary. On the other hand, $A_\varphi$ generates
``gauge" transformations acting on itself.  Indeed, let $Q(\eta) =
(k/2\pi)\int \eta_a A^a_\varphi$, it follows directly from (\ref{KM'})
that,
\begin{equation}
\delta A^a_\varphi = [A^a_\varphi, Q] = D_\varphi \eta^a.
\label{deltaA}
\end{equation}
However, here the interpretation is quite different because
(\ref{deltaA}) is not generated by the ``Gauss law" constraint $F=0$.
Instead, it is generated by $A_\varphi$ which is different from zero.
The symmetry (\ref{deltaA}) is a global --not gauge-- symmetry. This
means that configurations which differ by a transformation of the form
(\ref{deltaA}) are physically distinct. This is the origin of boundary
degrees of freedom in Chern-Simons theory.  

(ii) We have only computed the bracket between the values of the gauge
field $A$, not the group element $g$. This will be enough for our
purposes but we remark that the problem of computing the bracket of
$g(x)$ with itself leads to interesting constructions which involve
quantum groups. Another remarkable application of Chern-Simons theory
which we will not consider here is knot theory\cite{Witten89}.

(iii) The algebra (\ref{KM}) is known as affine, or Kac-Moody, $SU(2)$
algebra. This algebra is a non-Abelian generalization of the usual
Heisenberg algebra $[a_n,a_m]=n\delta_{n+m}$. Note that the last term
in (\ref{KM}) is precisely the algebra of three oscillators. The first
term couples them and, for example, alter the number of degrees of
freedom (degeneracy).  Unitary representations for (\ref{KM}) are well
understood (see, for example,
\cite{Gepner-W,Goddard-O,diFrancesco-MS}) and they exist provided $k$
is an integer.

(iv) Finally, an exercise for interested students: derive (\ref{KM})
starting from (\ref{pAA}) by fixing the gauge $A_r=0$, solving the
constraints $F=0$ and constructing the Dirac bracket. Note that the
constraint $F=0$ is a differential equation which on a manifold with
boundary will necessarily lead to an integration function. Identify
this function with $A_\varphi$ above.

\section{The Brown-Henneaux conformal symmetry}    

\label{Entropy}

As a final point, we briefly mention one the main application of the
affine algebra (\ref{KM}) to three-dimensional gravity. The content of
this section follows the original papers \cite{BH,CHvD} for the
derivation of the conformal algebra,
\cite{Polyakov90,Alekseev-S,Forgacs-WBFO} for the $SU(2)_k\rightarrow
$ Virasoro reduction, and \cite{Strominger97,Birmingham-SS,Carlip98-1}
for the statistical interpretation of the conformal algebra. See
\cite{Navarro-N,Baires,Twisted} for other aspects and further
developments.   

Let us show how does the Brown-Henneaux conformal algebra for anti-de
Sitter spacetimes is derived from (\ref{KM}).  We follow \cite{CHvD}.
(See \cite{B} for an alternative derivation based on a twisted
Sugawara construction, and \cite{BBCHO} for a supersymmetric
generalization.)  

It was pointed out in \cite{CHvD} that the full affine algebra does
not represent the dynamics of anti-de Sitter spacetimes. Indeed,
computing the metric associated to the boundary conditions invariant
under (\ref{KM}), one discoveres that they match the boundary
conditions found in \cite{BH} only if one imposes the additional
restrictions \cite{CHvD}
\begin{equation}
T^3_n=0, \ \ \ \  T^+_n=\delta^0_n, 
\label{red}
\end{equation}
on the affine generators ($T^\pm = T^1 \pm i T^2$).

These reductions conditions were first studied in \cite{Polyakov90} in
the context of two-dimensional gravity. It was shown in that reference
that the residual algebra leads to a Virasoro algebra with a central
charge $c=-6k$. Starting from (\ref{KM}) this result can be proved as
follows.  We regard (\ref{red}) as a set of second class constraints
to be imposed in the algebra (\ref{KM}). We then construct the
Dirac\footnote{For those not
familiar with the Dirac bracket formalism, see \cite{HT-book} for a
complete treatment. The idea is to find the Poisson bracket acting on
a system with constraints. For example, a free particle in three
dimensions with a canonical kinetic term $\int p_i \dot q^i$ has the
standard Poisson bracket structure. Suppose we decide to restrict the
movement of the particle according to $q^3=0,p_3=0$. The new Poisson
bracket is the same as before with the only modification that the
coordinates $q^3$ and $p_3$ are removed. There are cases, however, in
which the constraints are complicated functions of the canonical
variables and one can not remove the right coordinates just by
inspection. Let us consider a system with variables $z^a$ and a
Poisson bracket $[z^a,z^b]=J^{ab}$. Now, we impose the restrictions
$\phi_\alpha(z)=0$ such that det$C_{\alpha\beta}\neq 0$ where
$C_{\alpha\beta} = [\phi_\alpha,\phi_\beta]$. The Dirac bracket
$[~,~]^*= [~,~] - [~,\chi_\alpha]C^{\alpha\beta}[\chi_\beta,~]$ is
antisymmetric, satisfies the Jacobi identity and is invariant under
the constraints, $[X,\phi_\alpha]^*=0$ for all $X$.}
bracket $[~,~]^*$ which is invariant under (\ref{red}):
$[T^3_n,X]^*=[T^+_n,X]^*=0$ for all $X$. The only
remaining component $T^-_n$ can be renamed as $L_n=(1/k)T^-_n$ and it
follows that, in the Dirac bracket, $L_n$ satisfies the Virasoro
algebra 
\begin{equation}
[L_n,L_m]^* = (n-m) L_{n+m} + {c \over 12} n(n^2-1)\delta_{n+m}
\label{Virasoro}
\end{equation}
with central charge $c=-6k$ (see \cite{Baires} for an explicit
calculation).  From the value of $k$ given in (\ref{k}) we find,
\begin{equation}
c= {3l \over 2G},
\label{c}
\end{equation}
which is the correct Brown-Henneaux central charge\cite{BH}. 

The $c=3l/2G$ Virasoro algebra was discovered in 1986 \cite{BH}.
However only recently\cite{Strominger97} it was pointed out that it
plays a central role in the understanding of quantum three-dimensional
black holes.  The idea is the following.

The first input is that the zero modes $L_0$ and $\bar L_0$ of the
Virasoro algebra are related to the mass and spin of anti-de Sitter
spacetime as \cite{BH},
\begin{equation}
Ml = L_0 + \bar L_0 -{c \over 12}, \ \ \ \ \ \ J= L_0 - \bar L_0.
\label{MJ}
\end{equation}
For the black hole\cite{BHTZ}, these two parameters are related to the
inner and
outer horizon via, 
\begin{equation}
Ml = {r_+^2 + r_-^2 \over 8Gl}, \ \ \ \ \ J = {2 r_+ r_- \over 8Gl}.
\label{r+-}
\end{equation}

The Virasoro algebra (\ref{Virasoro}) represents a symmetry of the
theory, just like the angular momentum algebra,
$[L_i,L_j]=i\varepsilon_{ijk}L_k$, is the symmetry algebra of a
rotational invariant Lagrangian. Suppose that the algebra
(\ref{Virasoro}) is the symmetry algebra associated to some conformal
field theory which is unitary ($L_0,\bar L_0 \geq 0$) and modular
invariant.  Modular invariance implies that the partition function, 
\begin{equation}
Z[\tau] = \mbox{Tr}\, e^{2\pi i \tau (L_0-c/24) - 
2\pi i \bar\tau (\bar L_0-c/24)}, 
\label{Ztau}
\end{equation}
satisfies
\begin{equation}
Z[\tau'] = Z[\tau], \ \ \ \ \ \tau' = {a\tau +b \over c\tau + d},  
\label{mod}
\end{equation}
for any $a,b,c,d\in Z$ and $ad-bc=1$. The parameter $\tau$ is the
modular parameter, or complex structure, of the torus on which the CFT
is defined.  We recall that the partition function (\ref{Ztau}) has a
precise interpretation in the black hole manifold. The Euclidean black
hole has the topology of a solid torus whose modular parameter is
given in (\ref{tau}) and $L_0+\bar L_0$ is the Hamiltonian of the
theory (up to an additive constant that we discuss below). This is
actually implicit in (\ref{MJ}). 

Since $Z[\tau]$ is modular invariant, we can evaluate $Z[-1/\tau]$ in
the limit Im$(\tau)\rightarrow 0$. Assuming $L_0,\bar L_0 \geq 0$ we
obtain,
\begin{equation}
Z[\tau] \sim \left| \exp \left( {2\pi i c \over 24 \tau}\right)
\right|^2 . 
\label{canonical}
\end{equation}
From (\ref{tau}), $\tau = i\beta/2\pi l$ (non-rotating case), and
(\ref{c}) we find exactly the semiclassical Gibbons-Hawking limit
(\ref{Z3}). (Exercise: generalize this to the rotating case.) This is
the canonical \cite{BHTZ,BBO,Maldacena-S,BM} version of the results
obtained in \cite{Strominger97}.  Note that the limit
Im$(\tau)\rightarrow 0$ corresponds to small $\beta$ and, according to
(\ref{T3}), large values of $M$. This is a characteristic of the three
dimensional black hole not shared by the Schwarzschild solution. The
temperature in three dimensions decreases with the mass, the specific
heat is positive and the canonical ensemble is well-defined.     

A microcanonical calculation follows by  writing the partition
function (\ref{Ztau}) in the form,
\begin{equation}
Z[\tau] = \sum_{L_0,\bar L_0} \rho(L_0,\bar L_0) 
 e^{2\pi i \tau (L_0-c/24) - 2\pi i \bar\tau (\bar L_0-c/24)}, 
\label{Zm}
\end{equation}
where $\rho(L_0,\bar L_0) $ is the number of states with eigenvalues
$L_0,\bar L_0$. Using the approximation (\ref{canonical}) in
(\ref{Zm}) one can extract the number of states $\rho(L_0,\bar L_0) $ 
by a contour integral obtaining,
\begin{equation}
\rho(L_0,\bar L_0) \sim e^{2\pi \sqrt{cL_0/6} + 2\pi c\sqrt{c\bar
L_0/6}}. 
\label{Rama}
\end{equation}
This is known as Cardy formula. It is amusing to check that using
(\ref{c}), (\ref{MJ}) and (\ref{r+-}), the associated entropy is
exactly equal to the Bekenstein-Hawking value $S= A/4G$ with $A=2\pi
r_+$ \cite{Strominger97}.  

Actually, the above calculation is true provided the black hole mass
is large enough: $Ml>>c/12$ (see (\ref{MJ})). The shift $-c/12$
appearing in (\ref{MJ}) (which should be written as $-c/24-c/24$) is
the source of a number of issues. For unitary theories, on which the
above calculation makes sense, it means that the mass spectrum is
$M\geq -c/12$ and thus not only black holes enter in the partition
function but also the conical singularities (particle solutions)
introduced in \cite{Deser-JT}.  Curiously when writing canonical
expressions for the Virasoro generators, either using the Liouville
approach \cite{CHvD} or the twisted Sugawara operator \cite{B}, one
finds $M\geq 0$. This looks fine because the entropy should be
associated to black holes spacetimes having horizons and not to the
particle solutions.  However, if one restricts the spectrum to
positive masses, then the saddle point approximation (\ref{canonical})
is not valid.   In summary, the CFT whose symmetry is generated by
(\ref{Virasoro}), and that we assumed existed, does not seem to be
related to general relativity. 

We shall end here. See \cite{Carlip98-1,Martinec98} for discussions on
this last point, \cite{Twisted} for a proposal to resolve this problem
within general relativity, and \cite{Martinec98,Giveon-KS,deBoer-ORT}
for the string theory side of it. 

\section*{Acknowledgments}

The author would like to thank the organizers of the VIII Mexican
School on Particles and Fields for the kind invitation to deliver
these lectures.  I would also like to thank S. Carlip, M. Henneaux,
M.Ortiz and A.Ritz for many conversations and correspondence which had
been very helpful to understand the ideas presented here.  Financial
support from CICYT (Spain) grant AEN-97-1680, and the Spanish
postdoctoral program of Ministerio de Educaci\'on y Ciencia is also
acknowledged.


\begin{thebibliography}{10}

\bibitem{Strominger-V} A. Strominger and C. Vafa,  Phys.
Lett. {\bf B379}, 99 (1996)  

\bibitem{Callan-M} C.G. Callan and J. Maldacena, 
Nucl.Phys. {\bf B472}, 591-610 (1996).

\bibitem{Rovelli} C. Rovelli, Phys. Rev. Lett {\bf 77}, 3288 (1996).

\bibitem{Ashtekar-BCK}
A. Ashtekar, J. Baez, A. Corichi, K. Krasnov, 
Phys. Rev. Lett {\bf 80}, 904 (1998).  

\bibitem{Achucarro-T} A. Ach\'ucarro and 
P.K. Townsend,  Phys. Lett. {\bf B180}, 89 (1986).

\bibitem{Witten88} E. Witten, Nucl. Phys. {\bf B 311}, 46
(1988).

\bibitem{BH} J.D. Brown and M. Henneaux, Commun.
Math. Phys. {\bf 104}, 207 (1986).

\bibitem{Deser-JT} S. Deser, R. Jackiw and G. 't Hooft,
 Ann. Phys. {\bf 152}, 220 (1984); S. Deser and R. Jackiw, 
 Ann. Phys. (NY), {\bf 153}, 405 (1984)

\bibitem{BHTZ} M. Ba\~nados, C. Teitelboim and J.Zanelli,
Phys. Rev. Lett. {\bf 69}, 1849 (1992); M. Ba\~nados, M. Henneaux, C.
Teitelboim and J.Zanelli,  Phys. Rev. {\bf D48}, 1506 (1993).

\bibitem{Strominger97} A. Strominger, 
High Energy Phys. {\bf 02} 009 (1998). 

\bibitem{Martinec98} E. Martinec, ``Conformal field theory, 
geometry, and entropy", hep-th/9809021 

\bibitem{Henningson-S} M. Henningson, K. Skenderis, 
J.High Energy Phys 9807, 023 (1998). 

\bibitem{Giveon-KS}
A. Giveon, D. Kutasov and N. Seiberg, 
``Comments on string theory on adS(3)", 
e-Print Archive: hep-th/9806194

\bibitem{deBoer-ORT} J. de Boer, H. Ooguri, H. Robins 
and J. Tannenhauser,
"String theory on adS$_3$, hep-th/9812046 

\bibitem{Skenderis99} K. Skenderis, 
``Black holes and branes in string theory", hep-th/9901050.  

\bibitem{Carlip-book} S. Carlip, {\it Quantum Gravity in 2+1
Dimensions}, Cambridge University Press (1998).

\bibitem{Gibbons-H} G. Gibbons and S.W. Hawking, 
Phys.Rev. {\bf D15}, 2752 (1977)

\bibitem{Gibbons77} G. Gibbons, Phys.Lett. {\bf A61}, 3 (1977). 

\bibitem{Maldacena97} J. Maldacena, 
Adv. Theor. Math. Phys. {\bf 2}, 231 (1998). 

\bibitem{Gubser-KP} 
S.S. Gubser, I.R. Klebanov, A.M. Polyakov,  
Phys. Lett. {\bf B428}, 105 (1998) 

\bibitem{Witten98} E. Witten, 
Adv. Theor. Math. Phys. {\bf 2} 253 (1998)

\bibitem{Nelson-R} J.E. Nelson and T. Regge, 
Nucl.Phys.{\bf B328}, 190 (1989); Phys.Lett. {\bf B272},
 213 (1991); Commun. Math. Phys. {\bf 141}, 211 (1991).

\bibitem{Birmingham-SS}
D. Birmingham, I. Sachs and A. Sen,
Phys.Lett. {\bf B424}, 275-280 (1998) 

\bibitem{Page93} D.N. Page, ``Black hole information", 
hep-th/9305040 

\bibitem{Hawking-E} S.W. Hawking and G.F.R. Ellis (1973), 
{\it The Large Scale Structure of Space-time}, 
Cambridge University Press.

\bibitem{Hawking74} S.W.Hawking, Nature (London) {\bf 248}, 30 (1974).

\bibitem{Horowitz92} G. Howowitz, ``The Dark Side of String 
Theory: Black Holes and Black Strings", hep-th/9210119.

\bibitem{BTZ3} M. Ba\~nados, C. Teitelboim and J.Zanelli,
               Phys. Rev. {\bf D49}, 975 (1994).

\bibitem{Birrell-D}  N.D. Birrell and P.C.W. Davies, 
``Quantum Fields in curved space time", 
Cambridge University Press (1982).

\bibitem{Wald84} R.M. Wald (1984), {\it General Relativity}
University Press, Chicago, USA. 

\bibitem{Horowitz96} G.T. Horowitz, ``The origin of 
black hole entropy in string theory", gr-qc/9604051. 

\bibitem{Regge-T} T. Regge and C. Teitelboim,  Ann.
Phys. (N.Y.) {\bf 88}, 286 (1974).

\bibitem{Brown-} J.D.Brown, E.A.Martinez and J.W.York,
             Phy. Rev. Lett. {\bf 66}, 2281 (1991) 

\bibitem{BTZ4} M. Ba\~nados, C. Teitelboim and J.Zanelli,
               Phys. Rev. Lett.  {\bf 72}, 957 (1994).

\bibitem{Navarro-N}
P. Navarro and J. Navarro-Salas, 
Phys. Lett. {\bf B439}, 262 (1998).

\bibitem{B}   M. Ba\~nados,  Phys. Rev. {\bf D52}, 5816 (1995).

\bibitem{BBO} M. Ba\~nados, T. Brotz and M. Ortiz, ``Boundary 
dynamics and the statistical mechanics of the 2+1 dimensional 
black hole", hep-th/9802076, To appear in Nucl.Phys.B

\bibitem{Maldacena-S} J. Maldacena, A. Strominger,
``adS(3) black holes and a stringy exclusion principle".
e-Print Archive: hep-th/9804085

\bibitem{Brotz-OR99} T. Brotz, M. Ortiz and A. Ritz,
``On Modular Invariance and 3D Gravitational Instantons", 
hep-th/9903222. 

\bibitem{Baires} M. Ba\~nados, 
``Three dimensional quantum geometry and black holes", 
hep-th/9901148

\bibitem{CHvD}
O. Coussaert, M. Henneaux, P. van Driel,  
Class.Quant.Grav. {\bf 12}, 2961 (1995). 

\bibitem{Carlip-T} S. Carlip, C. Teitelboim, Phys.Rev. {\bf
D51}, 622 (1995).

\bibitem{Carlip95} S. Carlip, Phys. Rev. {\bf D51}, 632 (1995)

\bibitem{Carlip97} S. Carlip, Phys. Rev. {\bf D55}, 878, (1997) 

\bibitem{Carlip98-1} S. Carlip, 
Class. Quant. Grav. {\bf 15} 3609 (1998) 

\bibitem{BBCHO} M. Ba\~nados, K. Bautier, O. Coussaert, 
M. Henneaux and M. Ortiz, Phys. Rev. {\bf D58} 085020 (1998) 

\bibitem{Gepner-W} D. Gepner and E. Witten, Nucl.Phys. {\bf
B278}, 493 (1986). 

\bibitem{diFrancesco-MS} P. Di Francesco, P. Mathieu, D. Senechal.
``Conformal field theory"
New York, USA: Springer (1997) 890 p.

\bibitem{Witten89} E. Witten,  Commun. Math. Phys. {\bf
121}, 351 (1989).

\bibitem{Moore-S} G. Moore and N. Seiberg, Phys.Lett.
{\bf B220}, 422 (1989); S. Elitzur, G. Moore, A. Schwimmer and
N. Seiberg, Nucl. Phys. {\bf B326}, 108 (1989)

\bibitem{Goddard-KO} P. Goddard, A. Kent and  D. Olive,
Comm. Math.Phys. {\bf 103}, 105, (1986).

\bibitem{BM}  M. Ba\~nados, F. M\'endez
Phys. Rev. {\bf D58} 104014 (1998) 

\bibitem{HT-book} 
M. Henneaux and C. Teitelboim, {\it Quantization of Gauge Systems}
(Princeton University Press, Princeton, 1992).

\bibitem{Witten84} E. Witten, Commun. Math. Phys. {\bf 92}, 
455 (1984).

\bibitem{Polyakov90} A.M. Polyakov,
   Int. J. Mod. Phys. {\bf A5} (1990) 833.

\bibitem{Alekseev-S} 
A. Alekseev and S. Shatashvili, Nucl. Phys. {\bf B323}, 719 (1989).

\bibitem{Forgacs-WBFO} P.  Forg\'acs, A. Wipf, J. Balog, 
L. Feh\'er and L. O'Raifeartaigh, 
      Phys. Lett. {\bf 227 B} (1989) 214.

\bibitem{Goddard-O} P. Goddard and D. Olive,
 Int. Journ. Mod. Phys. {\bf A1}, 303, (1986).

\bibitem{Twisted} M. Ba\~nados, ``Twisted sectors in three
dimensional gravity", hep-th/9903178

\end{thebibliography}
\end{document}